\documentclass[iop]{emulateapj}
\usepackage{graphicx}
\usepackage{url}
\usepackage{apjfonts}

\usepackage{tikz}
\usetikzlibrary{shapes.geometric, arrows}
\tikzstyle{startstop} = [rectangle, rounded corners, minimum width=3cm, minimum height=1cm,text centered, draw=black, fill=red!30]
\tikzstyle{io} = [trapezium, trapezium left angle=70, trapezium right angle=110, minimum width=3cm, minimum height=1cm, text centered, draw=black, fill=blue!30]
\tikzstyle{process} = [rectangle, minimum width=1cm, minimum height=1cm, text centered, draw=black, fill=orange!30]
\tikzstyle{decision} = [diamond, minimum width=3cm, minimum height=1cm, text centered, draw=black, fill=green!30]
\tikzstyle{arrow} = [thick,->,>=stealth]

\shorttitle{Improving the Quality of {\it RXTE} PCA Spectra}
\shortauthors{Garc\'{\i}a \& et al.}

\begin{document}

\title{An Empirical Method for Improving the Quality of {\it RXTE} PCA Spectra}

\author{Javier Garc\'ia\altaffilmark{1}, Jeffrey~E.~McClintock\altaffilmark{1},
  James~F.~Steiner\altaffilmark{1}, Ronald~A.~Remillard\altaffilmark{2},
  Victoria~Grinberg\altaffilmark{2}}

\altaffiltext{1}{Harvard-Smithsonian Center for Astrophysics, 
  60 Garden St., Cambridge, MA 02138 USA; javier@head.cfa.harvard.edu,
  jem@cfa.harvard.edu, jsteiner@head.cfa.harvard.edu}

\altaffiltext{2}{MIT Kavli Institute for Astrophysics and Space Research, 
  MIT, 70 Vassar Street, Cambridge, MA 02139, USA; rr@space.mit.edu, 
  grinberg@space.mit.edu}

%==================================================================================
%

\begin{abstract}

We fitted all of the several hundred {\it RXTE} PCA spectra of the Crab
individually to a simple power-law model; the total number of counts in the
composite spectrum is $>10^9$. We then used the spectrum of residuals to derive
a calibration tool, called {\tt pcacorr}, that we apply to large samples of
spectra for GX~339--4, H1743--322, and XTE J1550--564. Application of the tool
improved the quality of all the fits, and the improvement is dramatic for
spectra with $\gtrsim10^7$ counts. The Crab residual spectrum is somewhat
different for each of the five PCA detectors, but it was relatively stable over
the course of the mission. We recommend that {\tt pcacorr} be routinely applied
to spectra with $\gtrsim10^6$ counts and that one include a systematic error of
0.1\%, rather than the 0.5--1\% value that has customarily been used. We
expect that application of the tool will result in an increase in sensitivity
of the PCA to faint spectral features by up to an order of magnitude.

\end{abstract}


%
%==================================================================================
%
%
%
\section{Introduction}\label{sec:intro}

The Rossi X-ray Timing Explorer ({\it RXTE}), launched on 30 December
1995 and decommissioned on 4 January 2012, observed hundreds of X-ray
sources during its 16 year lifetime. {\it RXTE} carried three
instruments: The All Sky Monitor \citep[ASM;][]{lev96}, consisting of
three scanning detectors that surveyed essentially the full sky during
each orbit ($\sim 93$~min); the High Energy X-ray Timing Experiment
\citep[HEXTE;][]{rot98}, two clusters of phoswich scintillation
detectors sensitive over the energy range 15--250~keV; and the
Proportional Counter Array (PCA), a set of five xenon-gas proportional
counters with a total effective area of 6500~cm$^2$ sensitive over the
2--60~keV energy range.

The PCA was the principal instrument aboard {\it RXTE}. Over the course
of the mission, the PCA made $\approx 110,000$ pointed observations,
each with a typical duration of a few thousand seconds; the total
effective observation time for the mission was $\approx295$~Msec.
This paper describes and makes
readily available a new methodology for improving the quality of PCA
spectra. Specifically, we recalibrate the detector response using a
mission-averaged spectrum of the Crab Nebula. 

The PCA was comprised of five nearly-identical Proportional Counter
Units (PCUs), each with an effective area of 1600~cm$^2$. The
PCA offered superb timing capability ($\Delta t \approx 1~\mu$s) and
modest spectral resolution ($\sim 18\%$ at 6~keV). Moreover, the
stability and predictability of the relatively low background (2 mCrab)
allowed the PCA to explore the behaviors of a great variety of X-ray
sources \citep[e.g.,][]{kaa04,swa99,swa06}. Details on the technical
specifications, performance, and calibration of the PCA are available in
\cite{gla94,zha93,jah96,jah06,sha12}.

\cite{jah06} describe the response of the PCA from the beginning of the
mission until 2004. They showed that for many observations the energy
calibration of the instrument is limited by systematic errors below 10
keV. Specifically, they reported that the unmodeled variations in the
instrumental background are less than 2\% of the {\em observed}
background below 10~keV and less than 1\% in the 10--20~keV region.
They therefore advocated a general and conservative approach of
including an allowance for systematic error. Thereafter, it became
standard practice in reducing/analyzing PCA spectra to add in each pulse
height channel a systematic error of 0.5--1\% in quadrature with the
statistical error to mask the uncertainties in the model of the detector
response. Assuming a competent spectral model of a source in question,
this procedure allows one to achieve good fits.  

%values of reduced $\chi^2\sim1$.

The {\it RXTE} team later revised the PCA calibration, producing an
updated physical model for the instrumental response. These efforts are
fully described in \cite{sha12}, including the latest version of the
response generator ({\tt pcarmf} v11.7) and the channel-to-energy
conversion table ({\tt e05v04}). The new calibration, which was tested
on the whole archive of observations for the Crab, showed a great
improvement in performance with respect to earlier calibrations.
Concerning systematic error, \cite{sha12} recommended: ``For most
observations, the systematic error of 0.5\% is sufficient, while for
extreme cases it can be raised to 1.5\%"\footnote{For a discussion of how
systematic error affects confidence intervals, see \cite{wil06},
and for a thoughtful analysis on treating systematic error, including
practical recommendations, see \cite{han11}.}.

In this paper we go one step further in improving the calibration of the
PCA. We combine several hundred spectra of the Crab accumulated over the
lifetime of the mission to produce a single spectrum of residuals
(data/model), which we describe in Section~\ref{sec:fits} and refer to
throughout as a ``ratio spectrum.'' This ratio spectrum has the extreme
statistical precision expected for an analysis based on $>10^9$ source
counts. We show that in fitting any PCA spectrum with $\gtrsim10^6$
counts one can significantly reduce the effects of systematic errors due
to uncertainties in the standard detector response files by simply
dividing the observed counts by our ratio spectrum.

In Section~\ref{sec:fits} we compute a ratio spectrum, and in
Section~\ref{sec:corr} we obtain, via an iterative procedure, a sequence
of ratio spectra that quickly converges to our final product, which we
refer to as a ``correction curve.'' We compute a correction curve
for each individual PCU for each of the two principal gain intervals.
Throughout, we refer to the complete suite of correction spectra, plus a
Python script for correcting any PCA spectrum of interest, as the tool
{\tt pcacorr} (see Section~\ref{sec:disc}). We demonstrate the
performance of {\tt pcacorr} by fitting selected spectra of GX 339--4,
H1743--322, and XTE J1550--564 (Section~\ref{sec:newcor}), as well as
spectra of the Crab itself (Section~\ref{sec:crabcheck}).

The tool {\tt pcacorr} allows one to correct a PCA spectrum of a bright
source of interest and achieve an acceptable fit to a model using a
significantly lower level of systematic error. Our analysis indicates
that the level of required systematics after the correction is applied
is roughly $0.1\%$, rather than the standard prescription of 0.5--1\%.
Thus, application of the tool greatly increases the detection
sensitivity of the PCA. The benefits are obvious. For example, key
physical parameters may be more
accurately and precisely determined. Additionally, and despite the limited energy
resolution of the PCA, the greater sensitivity will likely enable the
detection of subtle and previously undiscovered spectral features, such
as photoelectric edges in disk-photospheric spectra \citep{kub10}, or
absorption lines in disk-wind spectra \citep[e.g.][]{nei12}.

The paper is organized as follows. In the following section we present
our analysis of the complete collection of archived PCA spectra of the
Crab. Our method for correcting any PCA spectrum of interest is
described in Sections~\ref{sec:gain}~and~\ref{sec:corr}. In
Section~\ref{sec:crabcheck} we first apply the method to individual Crab
spectra, and in Section~\ref{sec:newcor} we extensively apply the method
to many spectra of three black hole binaries. In
Sections~\ref{sec:allpcus} -- \ref{sec:layers} we examine in turn how
the correction differs among the five PCUs, the stability of the
correction over the mission lifetime, and how the correction depends on
the whether one considers a single detector layer or all three. We
close with a discussion and our conclusions in Section~\ref{sec:disc}.

%
%==================================================================================
%
\section{Fits to the Crab Data and the Creation of a Ratio
  Spectrum}\label{sec:fits}

We have fitted all the archived spectra of the Crab taken with the PCA.
The data for all five PCUs were reduced and background subtracted
following procedures described more fully in \cite{McClintock2006}.
Briefly, the event files and spectra were screened using the data
reduction tools from HEASOFT version 6.13. Data were taken in the
``Standard 2 mode,'' which delivers a PCA spectrum every 16~s. The bulk
of the paper is focused on spectra created by summing the counts
detected in all three gas layers of each PCU, although we have also
analyzed all the Crab data for the top layer only (see
Section~\ref{sec:layers}). Background spectra were obtained using
version 3.8 of the tool {\tt pcabackest}, along with a background model
{\tt pca\_bkgd\_cmvle\_eMv20111129.mdl}, which was provided by the PCA
team and yields improved background spectra for bright sources
(C. Markwardt, private communication). Background spectra were
subtracted from the total spectra using the tool {\tt mathpha}. The
response files were generated using the latest version of the response
generator {\tt pcarmf} (version 11.7) and energy-to-channel conversion
table (version {\it e05v04}) described in \cite{sha12}. We disregard
data collected during the first 108 days, which corresponds to
$\approx2$\% of the mission lifetime (Gain Epochs 1 and 2) for which the
calibration of the PCA is unreliable.

A normalization correction for detector dead time was not applied to the data presented here
(the typical correction factor was $\approx 1.05$, with extreme values of 1.019
and 1.069). Given our concern in this paper with systematic errors, we note
that energy-dependent deadtime effects can in principle occur for bright
sources because the source and background components are generally not
corrected for deadtime in a self-consistent manner. For the Crab Nebula, where
the deadtime is moderate and the source dominates the background out to 45~keV,
we find that energy-dependent deadtime effects do not appear to be important
and that they are unrelated to the sharp spectral features that are our concern
in this paper. However, as a general caution, we note that significant
systematic errors may arise in the analysis of soft broadband spectra of very
bright sources from an inconsistent treatment of deadtime effects.

To bring focus to the paper, we primarily discuss and present results
for the 554 spectra taken with PCU-2, the detector generally considered
to be the best-calibrated and the one most often in operation. The time
boundaries for the 554 archived spectra are associated with the time
intervals of PCA FITS files (FS4a*) for data collected in the Standard-2
mode. However, we have analyzed all the PCA spectra of the Crab, and
the {\tt pcacorr} tool can be applied to a spectrum obtained by any one
of the five PCUs. All of our data analysis is performed using the X-ray
spectral fitting package {\sc xspec} \citep{arn96}, version 12.7.1.
Thus, the criterion we rely on for goodness-of-fit in fitting models to
data is

\begin{equation}\label{eq:chi2}
\chi^2 = \sum_k^N{ \frac{[S(k) - M(k)]^2}{\sigma^2(k)}},
\end{equation}
where $S(k)$ and $M(k)$ are the source and model counts for channel $k$,
respectively, and the summation is over all $N$ channels. The squared
error for channel $k$ is

\begin{equation}\label{eq:sig}
\sigma^2(k) = S(k) + B(k) + \sigma^2_\mathrm{sys}(k),
\end{equation}
where $B(k)$ is the background counts and $\sigma^2_\mathrm{sys}(k)$ 
is the systematic error assumed for channel $k$. 
As a general rule, one expects that a good fit
has been achieved when $\chi^2\sim \nu$, where $\nu$ is the number of
degrees of freedom, given by the number of channels minus the number of
free fit parameters. Hence, it is customary to refer to the {\it
 reduced} $\chi_{\nu}^2 \equiv \chi^2/\nu$, which has an expected value
near unity if the data are well described by the model. While
Equation~\ref{eq:chi2} defines the goodness-of-fit statistic that is
widely used in X-ray astronomy, several definitions of chi-squared are
discussed in the literature that differ largely in how the errors in
Equation~\ref{eq:sig} are estimated (e.g., see Section 7.4 in
\citealt{fei12}).

%
%==================================================================================
%
\begin{figure*}
\begin{center}
\includegraphics[scale=0.65,angle=0]{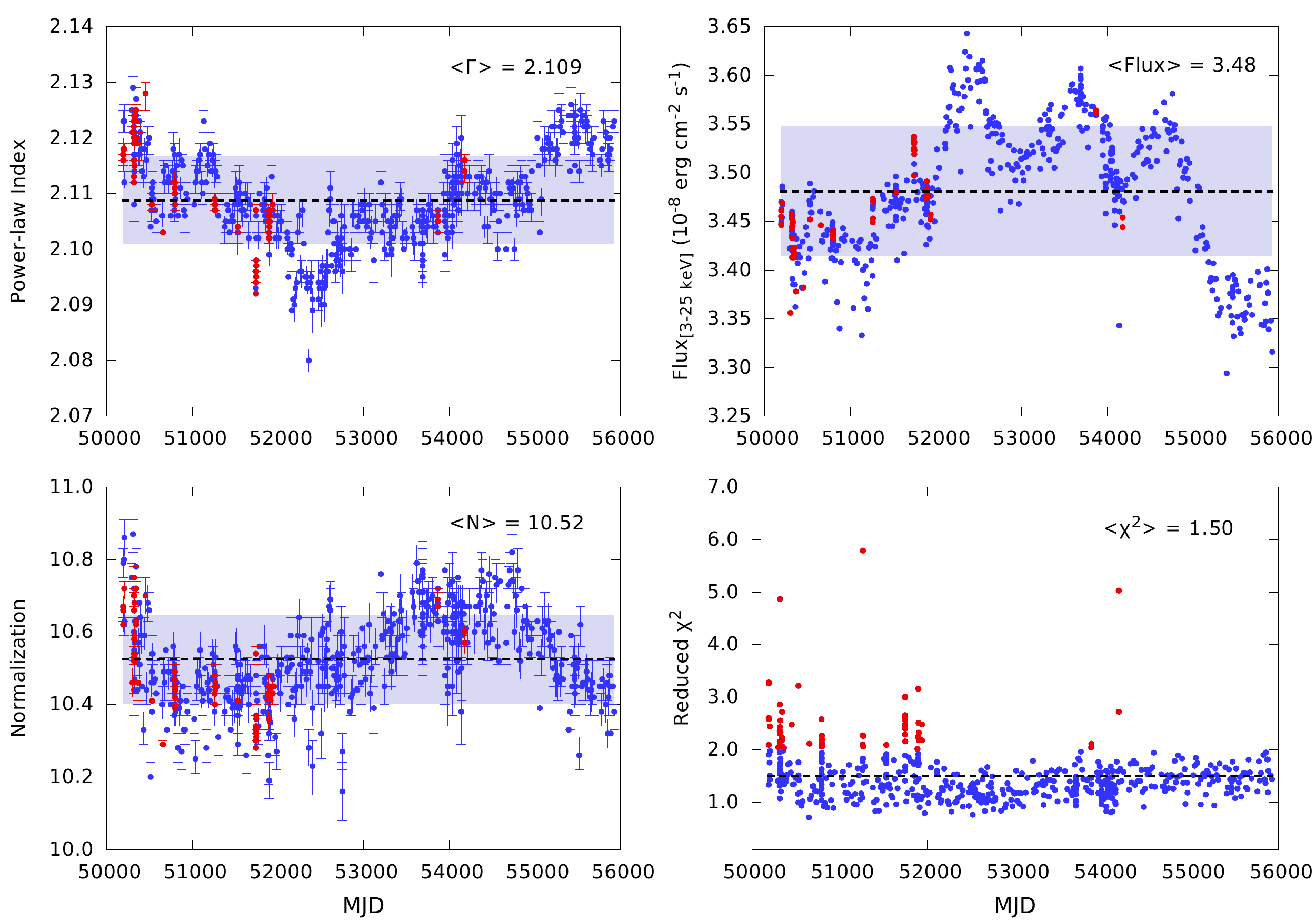}
\caption{Results of fitting 554 PCU-2 spectra of the Crab to an absorbed
power-law model with a fixed column depth of $N_{\rm H}=3.45\times
10^{21}$~cm$^{-2}$. Panels show photon index, normalization, flux, and
$\chi_{\nu}^2$ resulting from the individual fits. Average values are marked with
dashed lines and the shaded regions in the first three panels indicate their
standard deviations. Data points in red correspond to fits with $\chi_{\nu}^2 >
2$.}
\label{fig:plfits}
\end{center}
\end{figure*}
%==================================================================================

Because our goal is to assess systematic errors and improve the
calibration of the detectors, we do {\em not} follow the standard
practice of including an allowance for systematic errors in fitting
spectra; we consider only the errors due to counting statistics (i.e.,
$\sigma^2_\mathrm{sys}(k)=0$), except where otherwise specified.
Following \cite{sha12}, we ignored the
unreliable data in channels 1--4, and noticed energies up to 45
keV. Each Crab spectrum is fitted to an absorbed power-law using {\tt
 Tbabs*powerlaw} in {\sc xspec}, assuming a fixed hydrogen column
density of $N_\mathrm{H} = 3.45\times 10^{21}$~cm$^{-2}$. For the {\sc
 tbabs} model \citep{wil00}, we used the \cite{and89} set of solar
abundances and the \cite{bal92} cross sections. The resulting power-law
indices, normalization, flux, and $\chi_{\nu}^2$ for each fit are shown
in Figure~\ref{fig:plfits}. Our results are similar to those shown in
\cite{sha12}; specifically, the power-law index and power-law
normalization display the same patterns of variability. The mean
power-law index, normalization, flux, and reduced $\chi^2$ for our
complete sample are 2.11, 10.52, $3.48\times
10^8$~erg~cm$^{-2}$~s$^{-1}$, and 1.50, respectively. The power-law
index is consistent with that found by \cite{sha12}; however, our mean
normalization is somewhat lower, likely because we do not correct the
count rates for the effects of detector dead time, and our mean reduced
$\chi^2$ is slightly higher. The variations in flux follow closely those
previously reported by \cite{wil11} who find that this variability is
intrinsic to the Crab.

\begin{figure}
\begin{center}
\includegraphics[scale=0.5,angle=0]{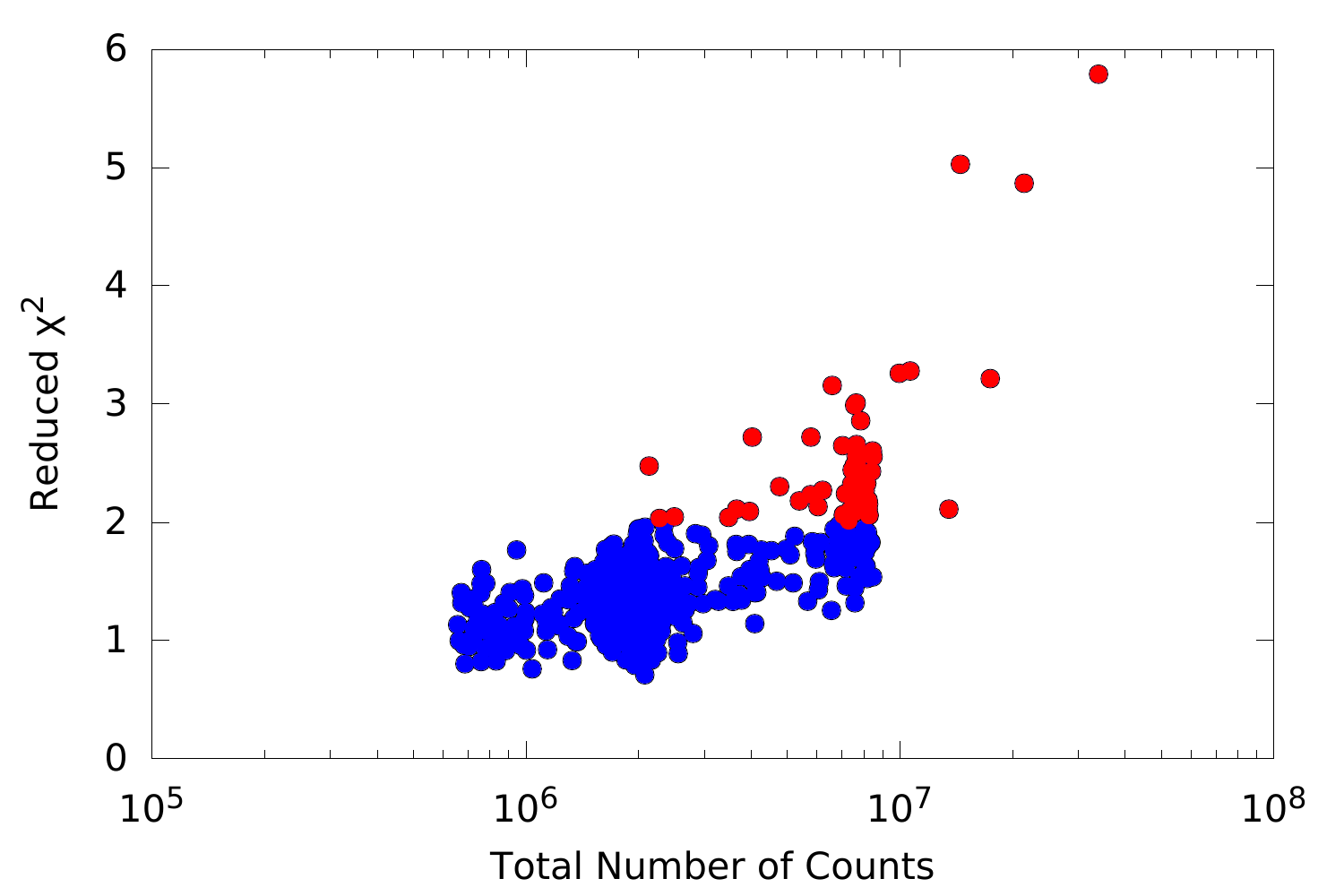}
\caption{$\chi_{\nu}^2$ as a function of total counts for the fits to
  all the 554 PCU-2 spectra of the Crab. Data points in red (which
  correspond to long exposures and many counts) indicate low-quality
  fits with $\chi_{\nu}^2>2$.}
\label{fig:chi2}
\end{center}
\end{figure}

About 7\% of the fits were relatively poor with $\chi_{\nu}^2 > 2$; the
corresponding data points in Figure~\ref{fig:plfits} are shown in red. Despite
the elevated values of $\chi_{\nu}^2$, the values of photon index and
normalization are typical of the full data sample. The poorer fit quality is
entirely due to longer exposure time, more counts, and hence the better
statistical quality of the data. Figure~\ref{fig:chi2} shows $\chi_{\nu}^2$
versus the total number of counts for each observation. Most of the
observations with $\chi_{\nu}^2 > 2$ (red symbols) have more than $5\times
10^6$ counts.

%==================================================================================
%
%=================================================================================

\begin{figure}
\begin{center}
\includegraphics[scale=0.5,angle=0]{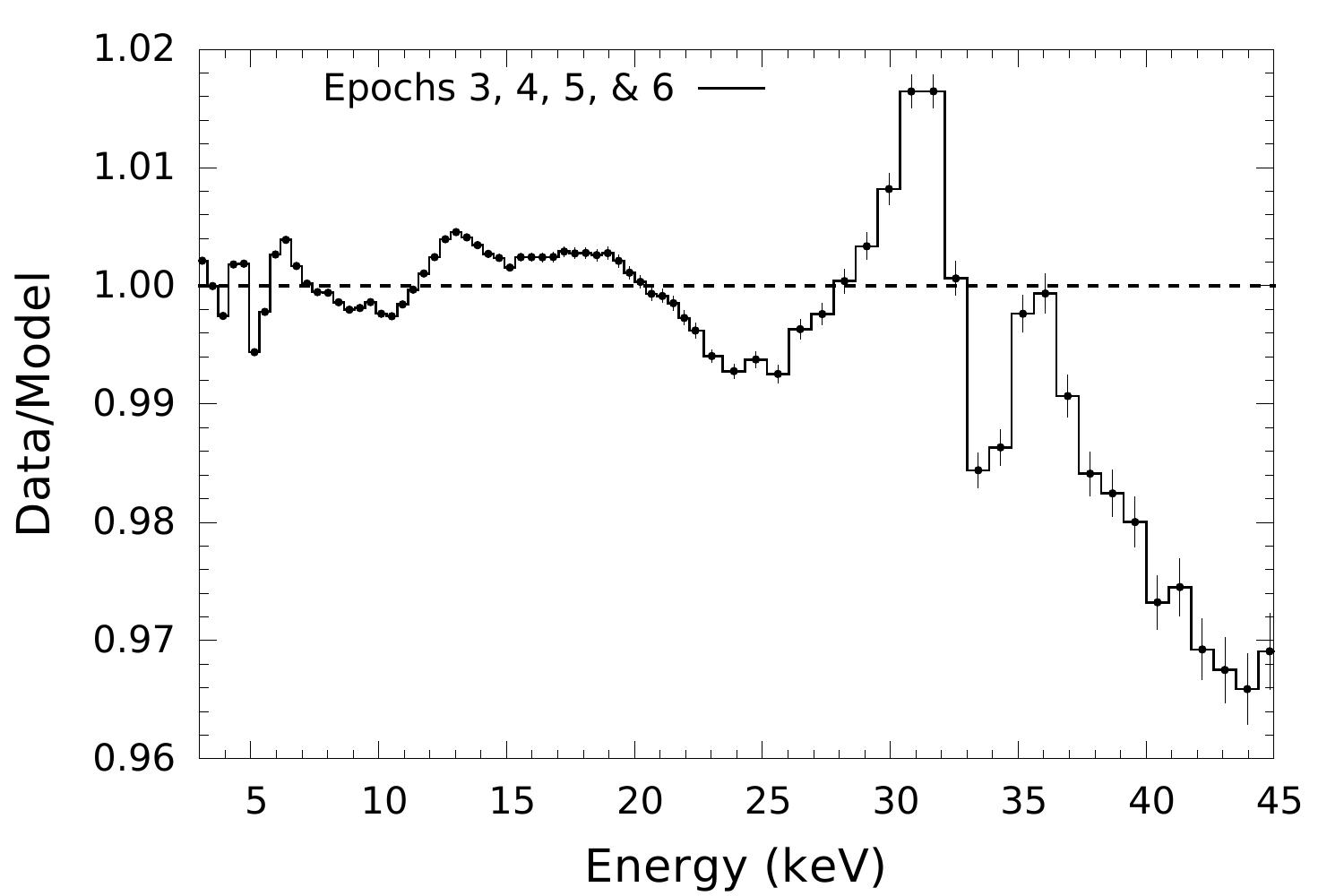}
\caption{Ratio spectrum obtained by combining all 554 PCU-2 Crab
  spectra. This summation of 746~ks of Crab data, corresponding to
  $2\times 10^9$ total counts, results in a ratio spectrum with
  extraordinary statistical precision.}
\label{fig:ratio}
\end{center}
\end{figure}

Figure~\ref{fig:ratio} shows a standard data-to-model ratio spectrum
produced by combining the data counts and the model counts for all 554
fitted Crab spectra. The total data counts $S$ from the source (i.e.,
excluding the background) and the model counts are simply obtained by
adding the net counts in each individual channel. A complication in
combining the various spectra is that the energy assigned to each
boundary is time-dependent, with discontinuous jumps at gain epoch
boundaries and smooth evolution during a given epoch\footnote{\url{
    http://heasarc.gsfc.nasa.gov/docs/xte/e-c\_table.html }}. We deal
with this problem by mapping all the individual spectra to the grid of
energies of one of the spectra (the reference spectrum), which can be
chosen at will; the ratio spectrum shown in Figure~\ref{fig:ratio} was
computed using the {\it RXTE} data set corresponding to obsID
60079-01-17-00.  The mapping is accomplished via linear interpolation.

%=================================================================================
%
%==================================================================================

The error bars displayed in Figure~\ref{fig:ratio} are given by the
ratio of the square root of the total data counts $D$ for each channel
to the model counts (Equation~\ref{eq:sig}). The combined ratio plot
shown in Figure~\ref{fig:ratio}, which was computed using all 554 PCU-2
Crab spectra, is equivalent to what one would obtain by analyzing a
single 746~ks Crab spectrum with $2\times 10^9$ counts. The
highly-significant features in the ratio spectrum
(Figure~\ref{fig:ratio}) represent departures from the pure power-law
spectrum expected for the Crab. By combining all the observations, we
are able to sensitively probe the $\sim1$\% irregularities in the
detector response, which have motivated observers over the years to
include a typical allowance of $\sim1$\% for systematic errors when
fitting PCA spectra (Section~\ref{sec:intro}). Assuming that the
synchrotron spectrum of the Crab is featureless and can be described by
an unbroken power law, the irregularities are due to imperfections in
the calibration of PCU-2. The ratio spectrum is fundamental to our
method of significantly decreasing the level of systematic error in the
data.

Possibly, the relatively sharp residual features shown in
Figure~\ref{fig:ratio} are unmodeled features associated respectively
with the Xe L-edges known to be present at 4.78, 5.10, and 5.45~keV, and
the Xe K-edge at 34.5~keV \citep{jah06}. Additionally, at energies above
$\sim30$~keV the spectrum generally trends downward, implying that the
best-fit power law over-predicts the observed spectrum by $\approx 3$\%
at 45~keV. It is not clear whether this effect is instrumental or
whether the Crab spectrum actually steepens. It is well-known that the
spectrum of the Crab is a composite of nebular and pulsar components and
that it breaks at $\sim100$~keV \citep[e.g.,][]{str79,jun89,yam11}.
Possibly, the Crab spectrum softens slightly in the PCA band, which
would imply a gradual softening of the spectrum over a wide range of
energies \citep{jou09} as opposed to an abrupt cutoff at $\sim100$~keV.
Alternatively, this downward trend above $\sim30$~keV may be related to
errors in the response files. That the effect is less significant 
in the ratio spectrum computed for the top xenon layer only 
(see Section~\ref{sec:layers}) supports this hypothesis.

%
%==================================================================================
%

\section{Dealing with Changing Gain}\label{sec:gain}

\begin{figure*}
\begin{center}
\includegraphics[scale=0.55,angle=0]{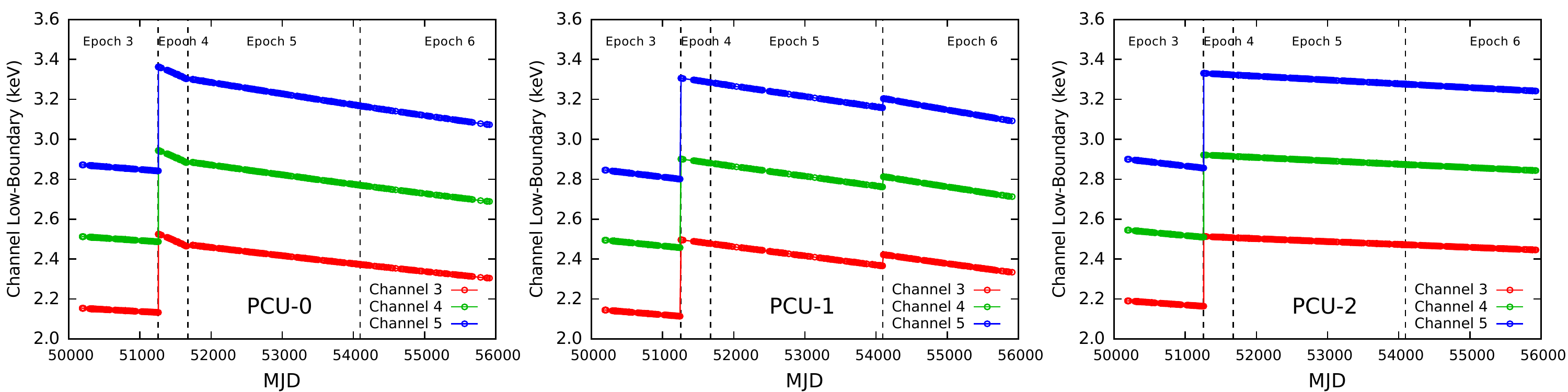}
\caption{Variation of the energy of the lower boundary of channels 3, 4,
  and 5 during the 16-years of the {\it RXTE} mission (excluding Gain
  Epochs 1 and 2). The data correspond to the hundreds of Crab spectra
  taken with PCU-0 (left), PCU-1 (middle), and with PCU-2 (right). The
  vertical lines show the transitions between gain epochs. There are
  clear and abrupt changes in the channel energies for all five PCUs in
  passing from Epoch 3 to Epoch 4, which occurred on MJD 51259. The
  transitions between Epoch 4 and Epoch 5, and between Epoch 5 and Epoch
  6, only affected PCU-0 and PCU-1, respectively; the effect on the
  channel boundaries of these detectors is small and we ignore them.}
\label{fig:channels}
\end{center}
\end{figure*}

As mentioned in Section~\ref{sec:fits} and illustrated in
Figure~\ref{fig:channels}, the gain of the PCUs changed continuously
throughout the mission, while discontinuous changes occurred at the
boundaries between gain epochs. The constantly changing gain of a detector
requires that each of the
Crab spectra are mapped to a common reference grid of energies. As we
show in Section~\ref{sec:temporal}, this procedure works well in
delivering a stable and useful ratio spectrum within a particular gain
epoch. However, as shown below, the ratio spectrum is not stable across
the major boundary between Gain Epoch 3 and Gain Epoch 4 (which we refer
to hereafter as Gain Epochs 4--6\footnote{We ignore the relatively
unimportant changes at the boundaries between Gain Epochs 4 and 5 and
Gain Epochs 5 and 6, which affected in each case only an individual
detector (see Figure~\ref{fig:channels}.)}) because the
channel-to-energy assignments changed quite significantly. The gain
discontinuity at this boundary is most problematic for the low-energy
channels ($E \lesssim 7$~keV) whose statistical precision is extreme.

%==================================================================================
%
%==================================================================================

\begin{figure}
\begin{center}
\includegraphics[scale=0.5,angle=0]{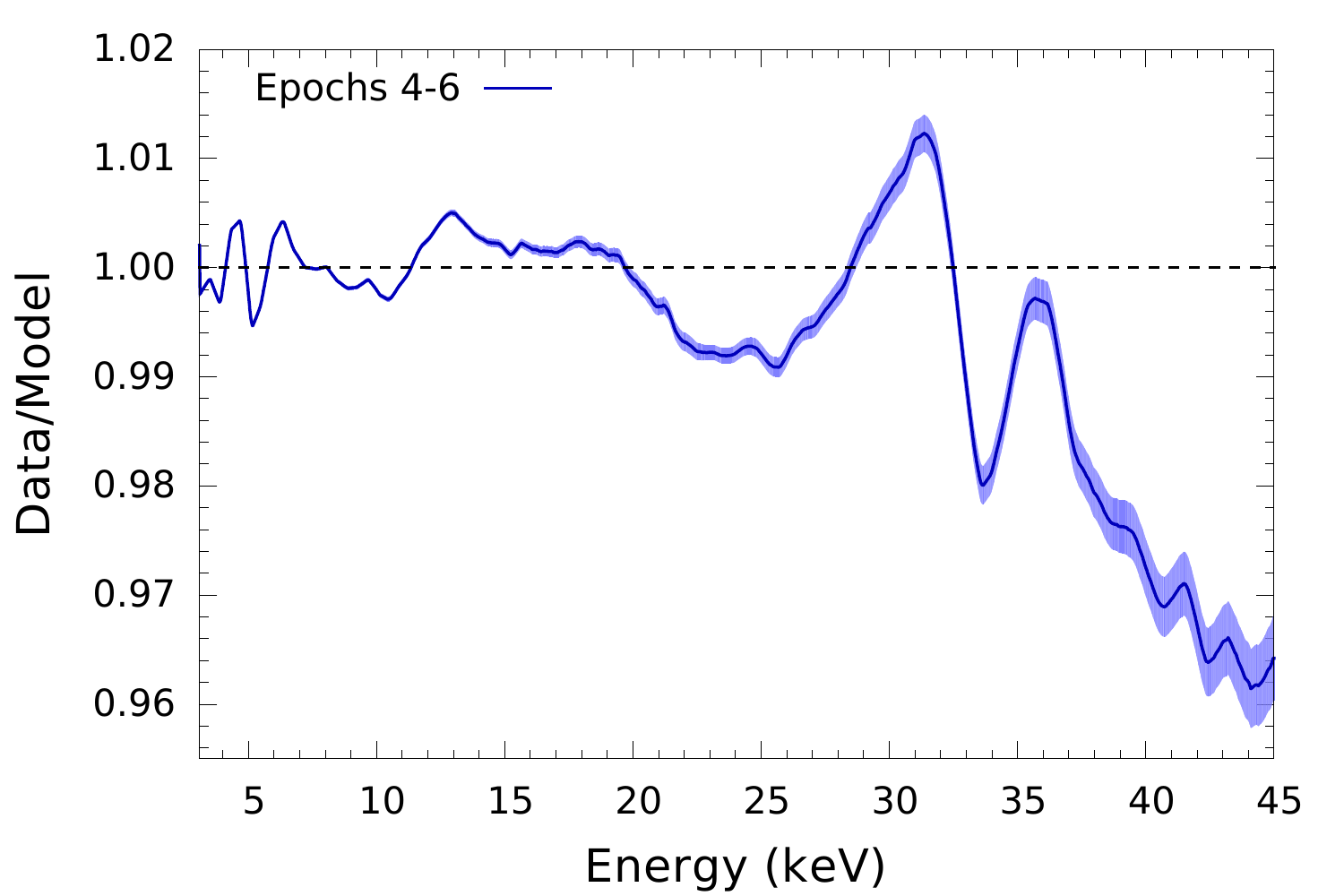}
\caption{Ratio spectrum comparable to the one shown in
  Figure~\ref{fig:ratio} but computed using a high-resolution grid and
  for Gain Epochs 4--6 only. The lighter shaded region shows the level of
  the statistical error.}
\label{fig:ratiohr}
\end{center}
\end{figure}

Instead of mapping all spectra to the grid of energies of one particular
spectrum, as we did in Section~\ref{sec:fits}, we now create a
synthetic, high-resolution reference energy grid \citep[similar to the $E_p$
space used in][]{jah06}. Compared to the
histogram (Figure~\ref{fig:ratio}), this new approach provides a ratio
spectrum that resolves finer details and reduces interpolation errors.
We choose a grid that encompasses all possible energy shifts in the
channel boundaries occurring over the mission lifetime. The grid of
$10^4$ bins is homogeneous and covers the energy range 1--150~keV. Each
individual spectrum is mapped to the grid by linear interpolation.
Figure~\ref{fig:ratiohr} shows the finely-gridded ratio spectrum
resulting from the combination of all PCU-2 Crab spectra for Gain Epoch
4--6.

%==================================================================================
%
%=================================================================================

\section{Correction Curves for the Two Principal Gain
Epochs}\label{sec:corr}

At this point one could adopt the ratio spectrum shown in
Figure~\ref{fig:ratiohr} as a final product to be used in correcting all
PCU-2 spectra obtained during Gain Epochs 4--6. However, via an
iterative procedure we obtain a final product of much higher quality for
use in correcting PCA spectra that we refer to hereafter as a
``correction curve.'' We now describe the iterative procedure and the
creation of the correction curve.

\begin{figure*}
\begin{center}
\includegraphics[scale=0.5,angle=0]{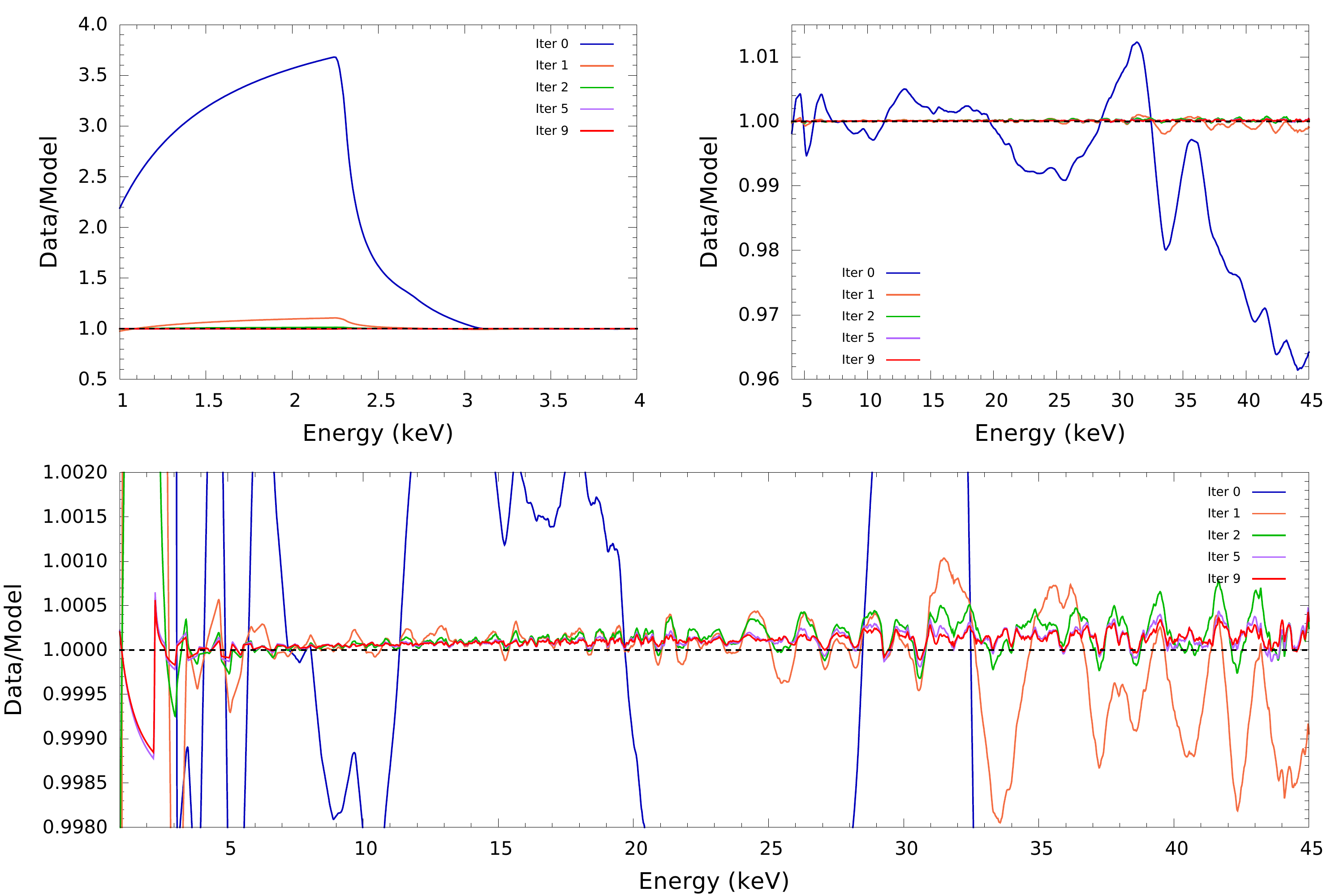}
\caption{Top panels: Low-energy (left) and high-energy (right) spectral
  regions showing several ratio spectra obtained by combining 417 PCU-2
  Crab spectra collected during Gain Epochs~4--6. The parent ratio
  spectrum labeled iter 0 (blue curve) is identical to that shown in
  Figure~\ref{fig:ratiohr}. The smaller-amplitude daughter spectra,
  which were computed using the iterative procedure described in the
  text, are sensitively displayed over the full 1--45 keV range in the
  lower panel. For most channels, nine iterations suffice to reduce the
  instrumental residuals to a level of $\sim 0.05\%$. }
\label{fig:iratio}
\end{center}
\end{figure*}

The blue curve in Figure~\ref{fig:iratio} (labeled iter 0) shows the
same ratio spectrum plotted in Figure~\ref{fig:ratiohr}, but with the
error bars suppressed for clarity. In the top two panels, this ratio
spectrum is plotted in two separate energy intervals, 1--4~keV and
4--45~keV, because of the very different scale of the residuals. We
perform the first iteration by dividing the counts in each of the 417
PCU-2 Crab spectra by the corresponding value given in the ratio
spectrum shown in blue and labeled ``iter 0'' in
Figure~\ref{fig:iratio}. In this way we produce a new ratio spectrum
plotted in orange (iter 1) with the instrumental residuals now reduced
to a level of $\sim0.2$\%. Correcting the Crab spectra a second time
using this ratio spectrum again further reduces the amplitudes of the
residual features. After nine iterations, the residuals have been
reduced to the extraordinary level of 0.05\% for most channels
(Figure~\ref{fig:iratio}). 

%=================================================================================
%
%=================================================================================

Although the residuals are largest for the low-energy channels, the
improvement is most dramatic for these channels because the data/model
ratio was initially $\sim3.5$ (top-left panel of
Figure~\ref{fig:iratio}). However, in computing the original ratio
spectrum (iter 0), we ignored channels 1--4 so that the correction at
low energies is based on an extrapolation. Although the correction
procedure dramatically reduces the residuals in these channels, the
calibration is highly uncertain and is a topic for future study. We
further note that, likewise, the extrapolation of the correction to
energies above the 45~keV limit of our fits, and over the entire energy
range of the PCA, also seems to perform well. The behavior of the
correction for channels 1--4 and for $E > 45$~keV is beyond the scope of
this paper and will not be discussed further.

\begin{figure}
\begin{center}
\begin{tikzpicture}[node distance=1.5cm]

\node (fit) [startstop] {Fit individual $S_j^i$ spectra};

\node (ratio) [process, below of=fit, align=center] {Produce combined ratio spectra \\
 $R^i = \sum_j S_j^i/\sum_j M_j^i$   };

\node (corr) [process, below of=ratio, align=center] {Correct individual spectra \\
 $S_j^{i+1} = S_j^i / R^i$ };

\node (dec) [decision, below of=corr, align=center, yshift=-1cm] {{\tt i=imax}?};

\node (cont) [process, left of=dec, xshift=-2cm] {{\tt i++}};

\node (end) [startstop, below of=dec, align=center, yshift=-1cm] {Final Correction Curve \\
 $C = \prod_i R^i$};
\node (end2) [startstop, below of=end, align=center] {Final Corrected Spectra \\
 $S_j^\dag = S_j^0/C$};

\draw [arrow] (fit) -- (ratio);
\draw [arrow] (ratio) -- (corr);
\draw [arrow] (corr) -- (dec);
\draw [arrow] (dec) -- node[anchor=south] {no}(cont);
\draw [arrow] (cont) |- (fit);
\draw [arrow] (dec) -- node[anchor=east] {yes}(end);
\draw [arrow] (end) -- (end2);

\end{tikzpicture}
\caption{Flow diagram showing the process employed in producing a
correction curve $C$ using a collection of Crab spectra.  The indices
  $i$ and $j$ denote the iteration number and the number of a particular
  spectrum, respectively.  (For simplicity, we omit the use of a third
  and obvious index for channel number). Ratio spectra $R^i$ computed
  for several iterations are shown in Figure~\ref{fig:iratio}, and the final
correction curve $C$ is shown in Figure~\ref{fig:ratio345}.
}
\label{fig:flow}
\end{center}
\end{figure}
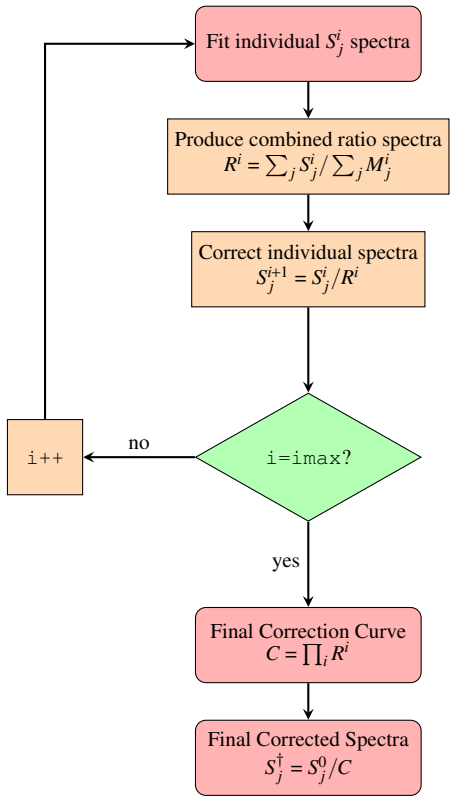

The flow diagram shown in Figure~\ref{fig:flow} summarizes the entire
process employed in producing the final correction curve.  The
individual Crab spectra in the set of observations $S_j^i$ are
identified by the index $j$, while the index $i$ is the iteration
number.  It is important to emphasize that the ratio spectra $R^i$ are
used at each iteration step $i$ to correct the current set of corrected
observations $S_j^i$; i.e., $R^i$ at each step is {\it not} used to
correct the original uncorrected data $S_j^0$.  Thus, as consecutive
iterations are performed, the magnitude of the correction decreases and
successive ratio spectra approach unity, as illustrated in Figure~\ref{fig:iratio}.
By visual inspection, we determined that after nine iterations the changes
in the ratio spectra were negligible (thus, imax=9).  Notice that
because the corrected set of spectra after $i$ iterations is
\begin{equation}
S_j^{i+1} = \frac{S_j^i}{R^i},
\end{equation}
then
\begin{equation}
S_j^{i+1} = \frac{S_j^{i-1}}{R^i R^{i-1}} 
         = \frac{S_j^0}{R^i R^{i-1}...R^0},
\end{equation}
and one can write a {\it final correction curve} as
\begin{equation}
C = \prod_i^\mathrm{imax} R^i.
\end{equation}
Denoting $S_j^\dag$ as the final set of corrected Crab spectra after imax
iterations, then the curve $C$ can be used to directly correct any set of 
source spectra $S_j^0$ simply by
\begin{equation}
S_j^\dag = \frac{S_j^0}{C}.
\end{equation}

{\it The final correction curve $C$, while derived using the Crab, can
  be applied to all sources}, i.e., it provides a detailed description
of the instrument response itself, and it is applicable to a PCA
spectrum $X_j^0$ of any object (including the Crab).  In fact, a suite
of 20 such universal corrections curves are computed, one for each
PCU, gain epoch, and detector configuration (see Section 10).

%=================================================================================
%
%
%===============================================================================

\begin{figure*}
\begin{center}
\includegraphics[scale=0.8,angle=0]{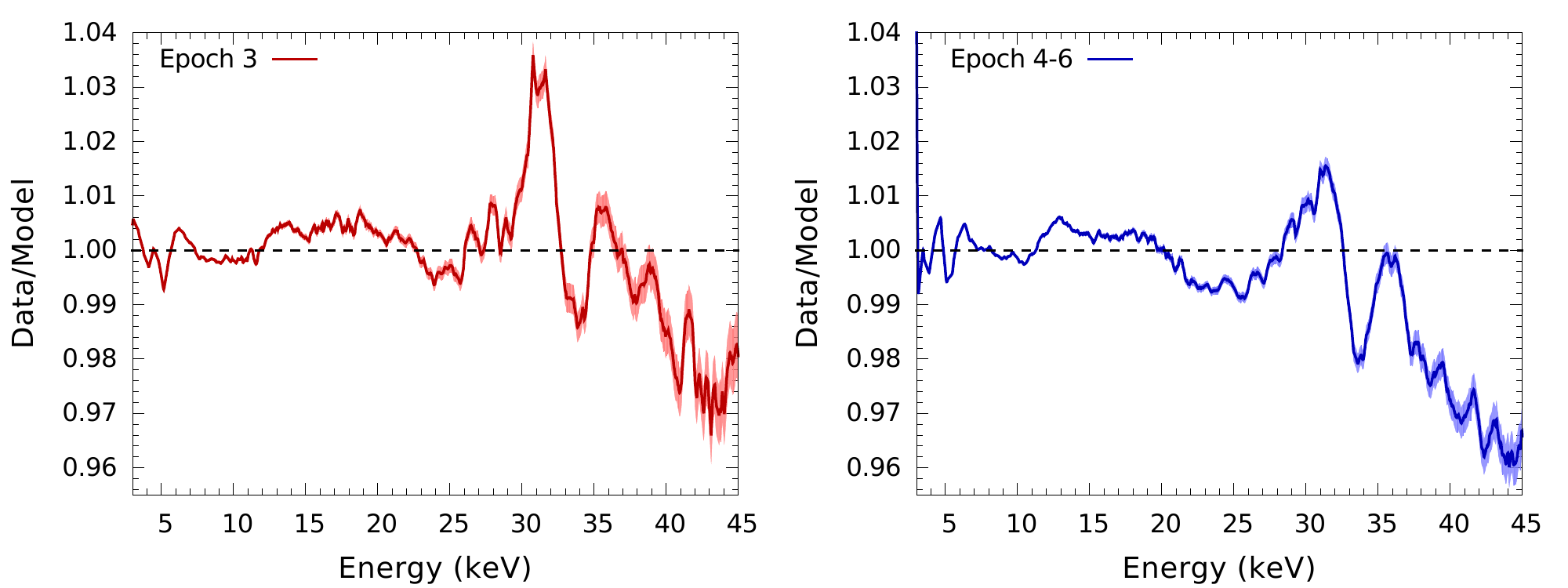}
\caption{Our adopted PCU-2 correction curves, one for each of the two principal
  gain periods, computed using a high-resolution energy grid after 9
  iterations (see Section~\ref{sec:corr}). Left: Computed for 137 PCU-2
  Crab spectra spectra taken during Gain Epoch 3; equivalent to a 262~ks
  observation with $6\times 10^8$~counts. Right: Computed for 417 PCU-2
  spectra taken during Gain Epochs 4--6; equivalent to a 484~ks
  observation with $10^9$~counts. The lighter shaded region bounding
  each curve shows the level of the statistical error.}
\label{fig:ratio345}
\end{center}
\end{figure*}

The final correction curve for PCU-2, which can be
used to correct the spectrum of any source observed during Gain Epoch
4--6, is shown in the right panel of Figure~\ref{fig:ratio345}. Because
of the abrupt change in the channel boundaries between Gain Epoch 3 and
Gain Epochs 4--6, we have produced a pair of correction curves for each
PCU; the correction curve for PCU-2 for Gain Epoch 3 is shown in the
left panel of Figure~\ref{fig:ratio345}. Meanwhile, we ignore the
boundaries between Gain Epochs 4 and 5, and between Gain Epochs 5 and 6,
because they are relevant only for PCU-0 and PCU-1 (which lost their
propane layers), and these discontinuities are weak and do not affect our
results.

%===============================================================================
%
%===============================================================================

The pair of correction curves shown in Figure~\ref{fig:ratio345} are of
central importance in the rest of the paper. While the curves are quite
dissimilar at energies $\lesssim7$~keV, they are qualitatively similar
at energies above $\sim7$~keV; e.g., both spectra show a prominent peak
near 30~keV and a downward trend above that energy. The Epoch~3 and
Epochs~4--6 correction curves (Figure~\ref{fig:ratio345}) were computed
using 137 and 417 Crab spectra, and the duration of each was
18\% (2.9~yrs) and 80\% (12.8~yrs) of the mission lifetime,
respectively. Meanwhile, we ignore the data for Gain Epochs 1 and 2
(2\% of the mission lifetime) because the calibration of the detectors
is uncertain.

% A pair of correction curves was computed for each PCU, one for Gain
% Epoch 3 and the other for Gain Epochs 4--6; the pair of curves for PCU-2
% is shown in Figure~\ref{fig:ratio345}. To correct a PCA object spectrum
% in question, one simply divides both the source counts and the errors by
% the correction curve for the appropriate PCU and gain epoch.  

Instead of
correcting the spectrum, we could have equivalently chosen to correct
the response file.  We chose to correct the spectrum because the makeup
of response files prepared by different observers can differ
significantly (e.g., the observer may or may not choose to include the
effective area in the response file). In the following two sections, we
illustrate the use of the {\tt pcacorr} tool by correcting PCU-2 spectra
of the Crab and three black holes.
%
%==================================================================================
%
\section{Applying {\tt pcacorr} to Individual Crab Spectra}\label{sec:crabcheck}

\begin{figure}
\begin{center}
\includegraphics[scale=0.5,angle=0]{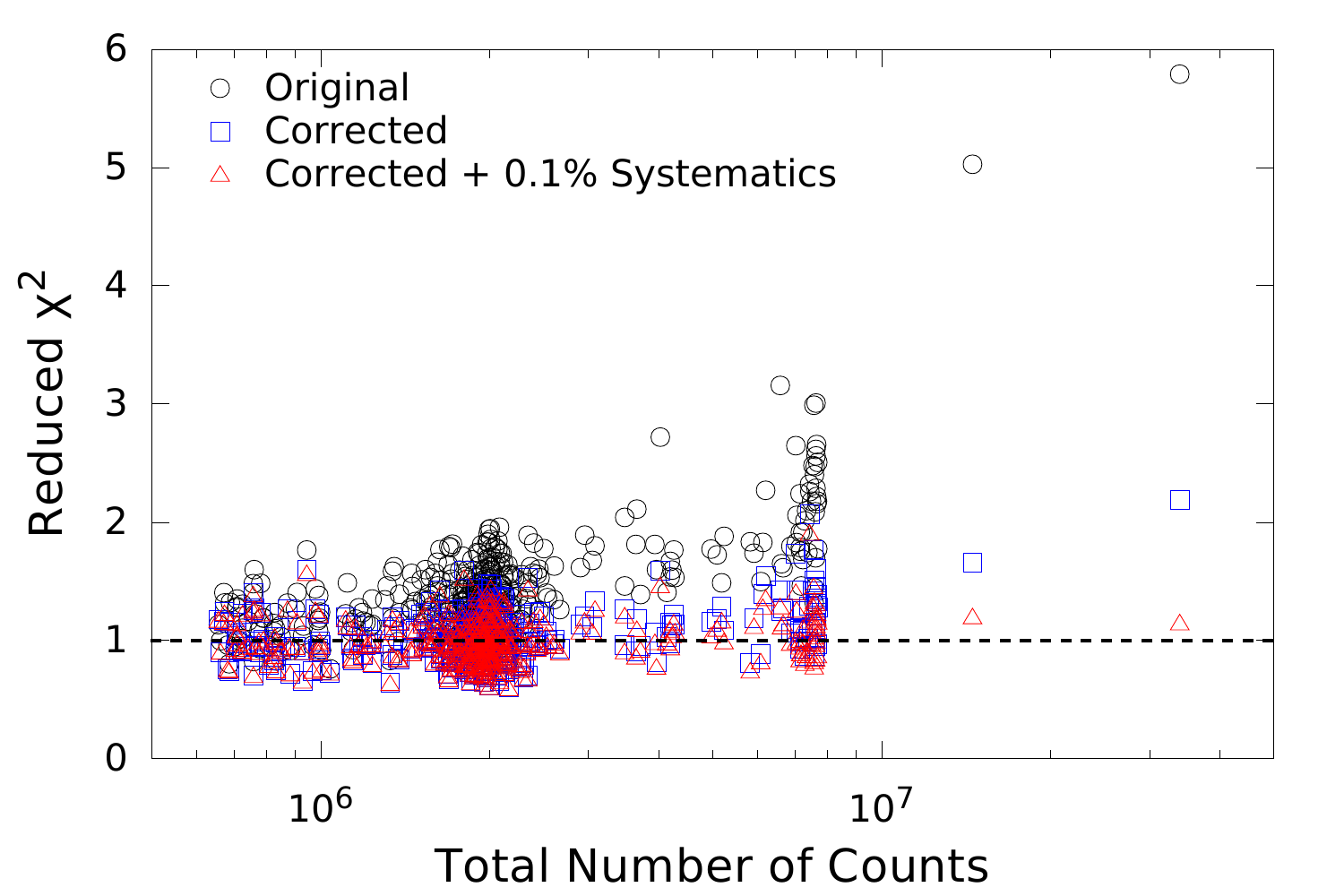}
\caption{ Goodness-of-fit ($\chi_{\nu}^2$) obtained for the 417 PCU-2
  Crab observations taken during Gain Epochs 4--6. The model is an
  absorbed power-law ({\tt TBabs*powerlaw}) with
  $N_\mathrm{H}=3.45\times10^{21}$~cm$^{-2}$. Black circles show
  $\chi_{\nu}^2$ for the original spectra (like in
  Figure~\ref{fig:chi2}), while the blue squares are for the spectra
  that have been divided by the correction curve shown in the right
  panel of Figure~\ref{fig:ratio345}. For the original and corrected
  data, the systematic uncertainty has been set to zero. Including a
  systematic error of 0.1\% is sufficient to bring $\chi_{\nu}^2$ to
  essentially unity for all corrected spectra (red triangles). }
\label{fig:chi2corr}
\end{center}
\end{figure}

We now correct the 417 PCU-2 spectra of the Crab collected during Gain
Epochs 4--6. For each Crab spectrum, we divide the counts in each
channel, as well as the error, by the corresponding value of the
correction curve shown in the right panel of Figure~\ref{fig:ratio345}.
We then fit both the corrected and uncorrected spectra to a simple
power-law model precisely as described in Section~\ref{sec:fits},
ignoring channels 1--4 and data above 45~keV. As before, and throughout
this paper (apart from two minor exceptions, which are noted in the
following paragraph and in the following section) we do not include any
allowance for systematic errors when fitting the data. The values of
$\chi_{\nu}^2$ for the fits to the corrected and uncorrected spectra are
compared in Figure~\ref{fig:chi2corr}.

%===============================================================================
%
%%==================================================================================

In all cases, the correction improves the quality of the fit
significantly, although the improvement is unremarkable for observations
with fewer than $\sim 10^6$ counts. The mean values of the broadband
model parameters (i.e., photon index, normalization, and flux) are
essentially unchanged by the correction.  The efficacy of the correction
increases markedly with the number of counts. Despite the excellent
performance of the {\tt pcacorr} tool, there is still an upward trend in
$\chi_{\nu}^2$ with the number of counts and fits to the spectra with
$\gtrsim10^7$ counts can be improved further. As one of the minor
exceptions mentioned above, we now make allowance for systematic error:
Including a systematic error of 0.1\%, we find that in almost all cases
$\chi_{\nu}^2$ decreases and that its values is now everywhere close to
unity.  The decrease is most obvious for the two longest exposure with
$> 10^7$ counts. We therefore suggest that a 0.1\% systematic error be
routinely included when fitting PCA data that has been corrected using
the {\tt pcacorr} tool.
%
%==================================================================================
%
\section{Testing {\tt pcacorr} on Spectra of Stellar-Mass Black Holes}\label{sec:newcor}

In applying the {\tt pcacorr} tool to the spectra of three well-studied
black hole binaries, GX~339--4, H1743--322, and XTE~J1550-564, we have
two goals: To establish whether the residual features evident in the
ratio spectrum (Figure~\ref{fig:ratio}) are indeed instrumental, and to
test the effectiveness of the correction in improving fits to
bright-source PCA spectra. As for the Crab spectra
(Section~\ref{sec:crabcheck}), both the counts and the error at each channel 
in the source spectrum are divided by the respective value of the
correction curve (while the errors in the latter are ignored).

\begin{figure}
\begin{center}
\includegraphics[scale=0.5,angle=0]{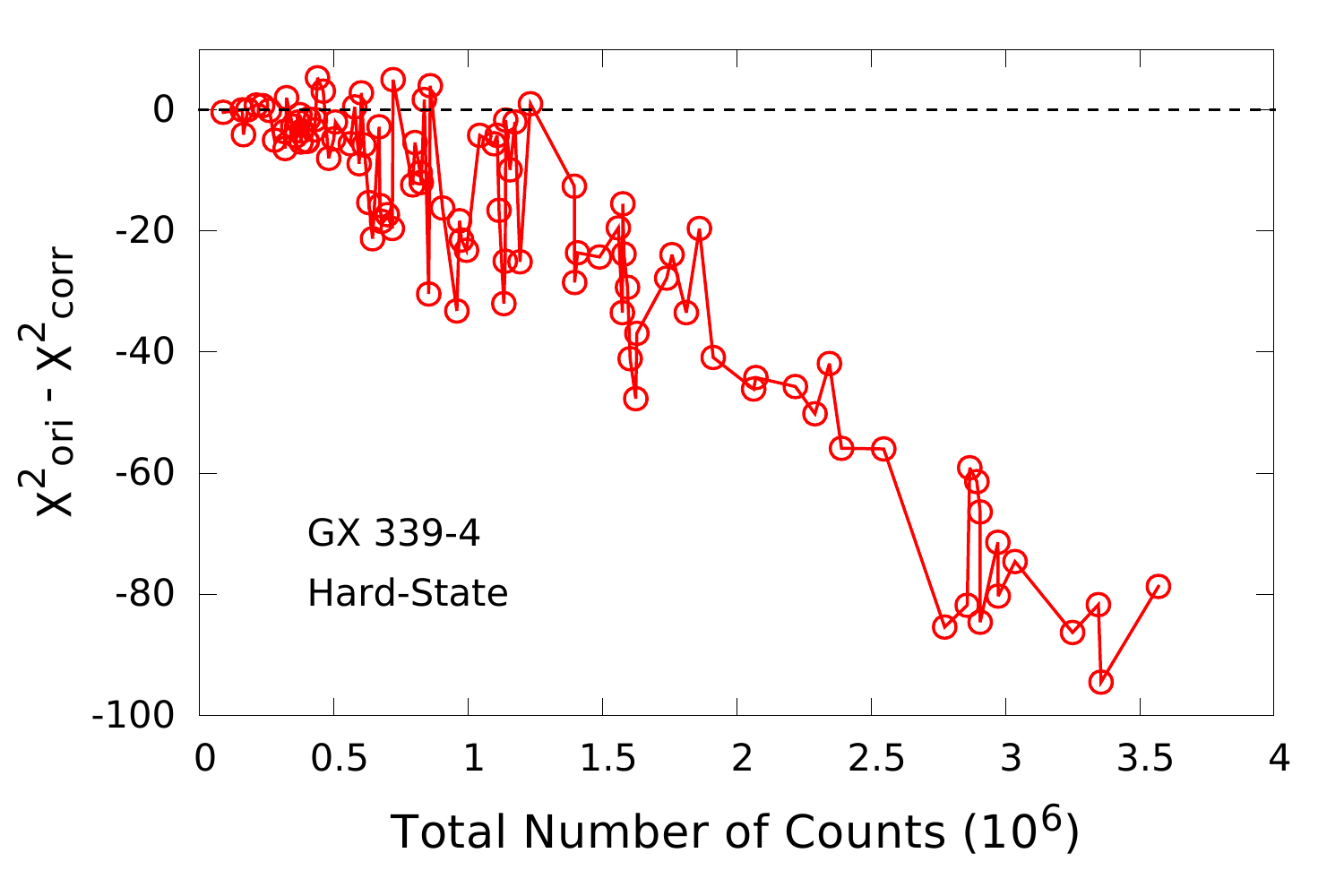}
\caption{Improvement in $\chi^2$ as a function of total counts for 100
  high-luminosity spectra of GX~339-4 in the hard state. The principal
  component in the model is an absorbed power law; a Gaussian and a
  smeared edge were included as cosmetic components. Plotted is the
  difference in $\chi^2$ between the original uncorrected spectra and
  their corrected counterparts. The data were corrected using the
  correction curve for Epochs~4--6 (right panel of
  Figure~\ref{fig:ratio345}). All spectra were taken with PCU-2. Data
  for channels 1--4 and for energies above 45~keV are
  ignored. Systematic errors are set to zero.}
\label{fig:chi-gx339}
\end{center}
\end{figure}

We first test the tool using high-luminosity PCU-2 spectra
of GX~339--4, specifically, the brightest 100 hard-state spectra
\citep[hardness $ > 0.75$; see][]{rem06}. These data were collected
during Gain Epoch 5, and we therefore use the ratio spectrum shown in
the right panel of Figure~\ref{fig:ratio345}. We ignore channels 1--4
and channels above 45~keV and set the systematic uncertainty to zero.
We again use a power-law model; however in this case, a smeared-edge
(constrained to 7--9~keV) and a Gaussian line (constrained to 6--7~keV)
are required to account for reflection features (which are not present
in the Crab): {\tt TBabs*smedge*(powerlaw+Gaussian)}. The model was
fitted both to the original spectra and to the corrected spectra; the
differences in $\chi^2$ are shown in Figure~\ref{fig:chi-gx339}. We find
consistently better fits for the corrected data, with $\Delta\chi^2$
steadily decreasing as the number of counts increases, as expected.

%==================================================================================
%
%==================================================================================

In this example and the two that follow, as a simple check on the
competency of the models we employ, we also fitted the same data sets
using the standard past practice of including a systematic error of 1\%;
as expected, we obtained satisfactory fits with $\chi_{\nu}^{2} \sim 1$.
Here, and in Section~\ref{sec:crabcheck} (where we derive our
recommendation that 0.1\% systematic error be routinely applied) are the
only places in this paper where we include any allowance for systematic
error.
 
\begin{figure}
\begin{center}
\includegraphics[scale=0.5,angle=0]{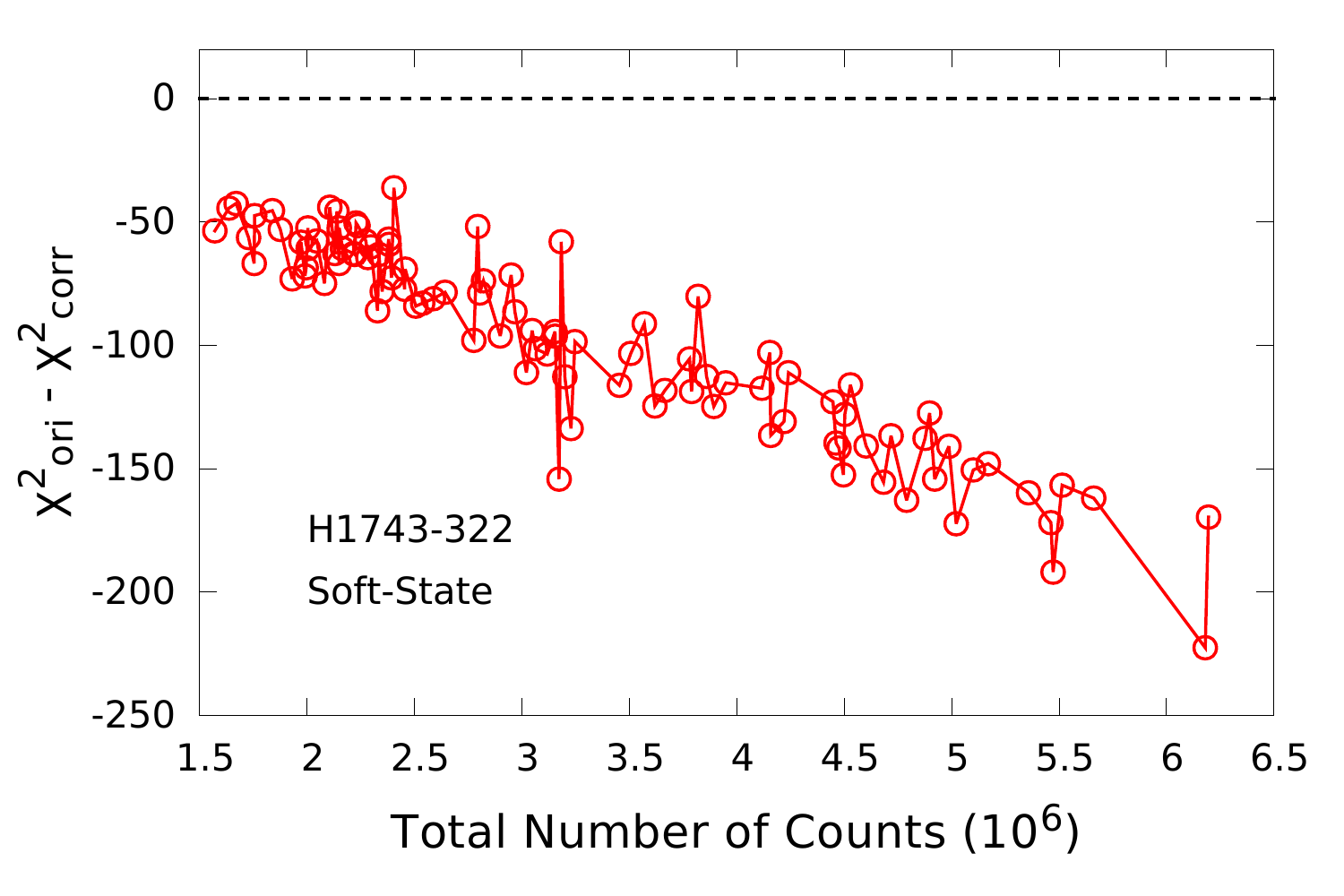}
\caption{Improvement in $\chi^2$ as a function of total counts for 100
high-luminosity spectra of H1743--322 in the soft state. The model
consists of a multi-temperature disk component as well as an absorbed
power-law component. The data were corrected using the correction
curve for Epochs~4--6 (right panel of Figure~\ref{fig:ratio345}).
Data for channels 1--4 and for energies above 45~keV are ignored.
Systematic errors are set to zero.}
\label{fig:chi-h1743}
\end{center}
\end{figure}

A similar test was performed on the 100 brightest soft-state PCU-2
spectra of H1743--322. In the soft (or thermal) state, the source
spectrum at low energies is dominated by the accretion-disk component.
Accordingly, our simple model includes a multi-temperature disk
component (while excluding the unnecessary reflection component): {\tt
TBabs*(powerlaw + diskbb)}. Our procedures are otherwise identical to
those used in fitting the spectra of GX 339--4 and the Crab.
Figure~\ref{fig:chi-h1743} shows for each spectrum the improvement in
$\chi^2$ achieved by applying the correction. As in the case of
GX~339--4, the benefit of applying the correction grows as the number of
counts increases. The results summarized in Figures~\ref{fig:chi-gx339}
and \ref{fig:chi-h1743} clearly demonstrate the efficacy of the
correction for both hard- and soft-state data, which indicates that the
tool has broad applicability.

%==================================================================================
%
%==================================================================================

\begin{figure}
\begin{center}
\includegraphics[scale=0.5,angle=0]{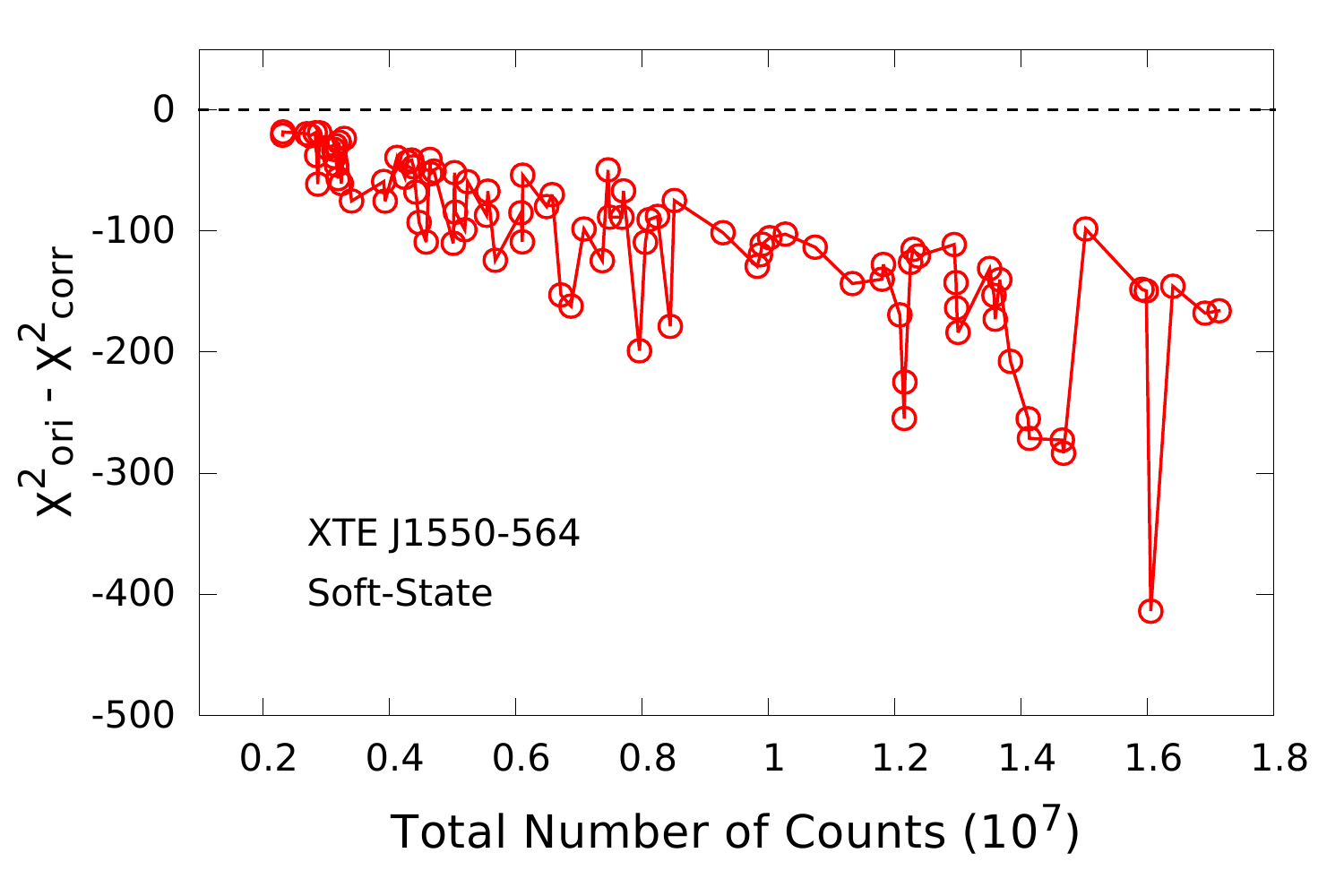}
\caption{Improvement in $\chi^2$ as a function of total counts for 94
high-luminosity spectra of XTE~J1550--564 in the soft state. The model
consists of a multi-temperature disk component and an absorbed power law
plus a smeared edge. The data were corrected using the correction
curve for Epoch~3 (left panel of Figure~\ref{fig:ratio345}). All
spectra were taken with PCU-2. Data for channels 1--4 and for energies
above 45~keV are ignored. Systematic errors are set to zero.}
\label{fig:chi-j1550}
\end{center}
\end{figure}

We have also tested the tool using Epoch~3 data, namely, 94 of the
brightest soft-state spectra of XTE J1550--564. In this case, the
relevant correction curve is the one shown in the left panel of
Figure~\ref{fig:ratio345}. The original and corrected data are fitted
with the same model for the continuum used in fitting the spectra of
H1743--322, except in this case we added a smeared Fe edge constrained
to 7--9~keV: {\tt Tbabs*smedge(powerlaw+diskbb)}. As shown in
Figure~\ref{fig:chi-j1550}, the results closely mirror those obtained
for GX~339--4 and H1743--322.

%==================================================================================
%
%==================================================================================

Finally, we tested the correction using an advanced reflection model
applied to a bright subset of the GX~339--4 hard-state spectra discussed
above. Specifically, we selected 23 spectra with normalized PCU-2 count
rates between 1000 and 1100 cts~s$^{-1}$, which corresponds to an
intensity of $\sim 0.4$ Crab. These spectra were obtained from consecutive
observations performed between 2002-04-20 and 2002-04-30 (MJD 52384.126
to 52394.444). The model consists of an absorbed power
law with exponential cutoff and a blurred X-ray reflection component. In
this instance, we used the latest version of the relativistic reflection
model {\sc relxill} \citep{gar14a}. The model is simply expressed as
{\tt Tbabs*relxill}, where {\sc relxill} already includes the power-law
component. Once again, a significant improvement is
consistently obtained by applying the correction.

In considering how the parameters of our physically well-grounded
reflection model are affected by making the correction, we have fitted
the same 23 spectra simultaneously with all the parameters tied except
for a normalization constant\footnote{This procedure is well motivated
  because the data are selected to have the same intensity and spectral
  hardness.}. The correction significantly improved the composite fit,
$\Delta \chi^2=313.04$; with no correction we obtained
$\chi_{\nu}^2=1.91$, while with the correction we obtained
$\chi_{\nu}^2=1.71$ (for 1605 degrees of freedom in both cases). As
always, we do not include any allowance for systematic error, which
contributes to the large values of $\chi_{\nu}^2$. Our purpose here is
not to perform a detailed analysis of these spectra, but rather to
determine if the correction significantly affects the parameters of this
detailed model.
%The parameters for these two fits are compared in Table~\ref{tab:relxill}.

Despite the large improvement in $\chi^2$, the model parameters are only
slightly affected. In fact, the parameters that describe the continuum
(e.g., photon index, ionization parameter, reflection fraction and
normalization) are unaffected by the correction. The only parameters
that are modestly affected are those informed by the reflection
component, i.e., the inclination angle, inner-radius of the accretion
disk, and Fe abundance. These are physically interesting model parameters
which control the strength and shape of both the Fe K emission line at 
$\sim 6.4$~keV and the Fe K edge at $\sim 7$~keV. 
This is a particularly important result
in the context of measuring the spins of black holes by the Fe-line
method, which relies on accurate modeling of the profile of the Fe K
emission line \citep[e.g.,][]{bre13b}.

%
%==================================================================================
%
\section{Comparing the Correction Curves for the Five PCUs}\label{sec:allpcus}

\begin{figure}
\begin{center}
\includegraphics[scale=0.5,angle=0]{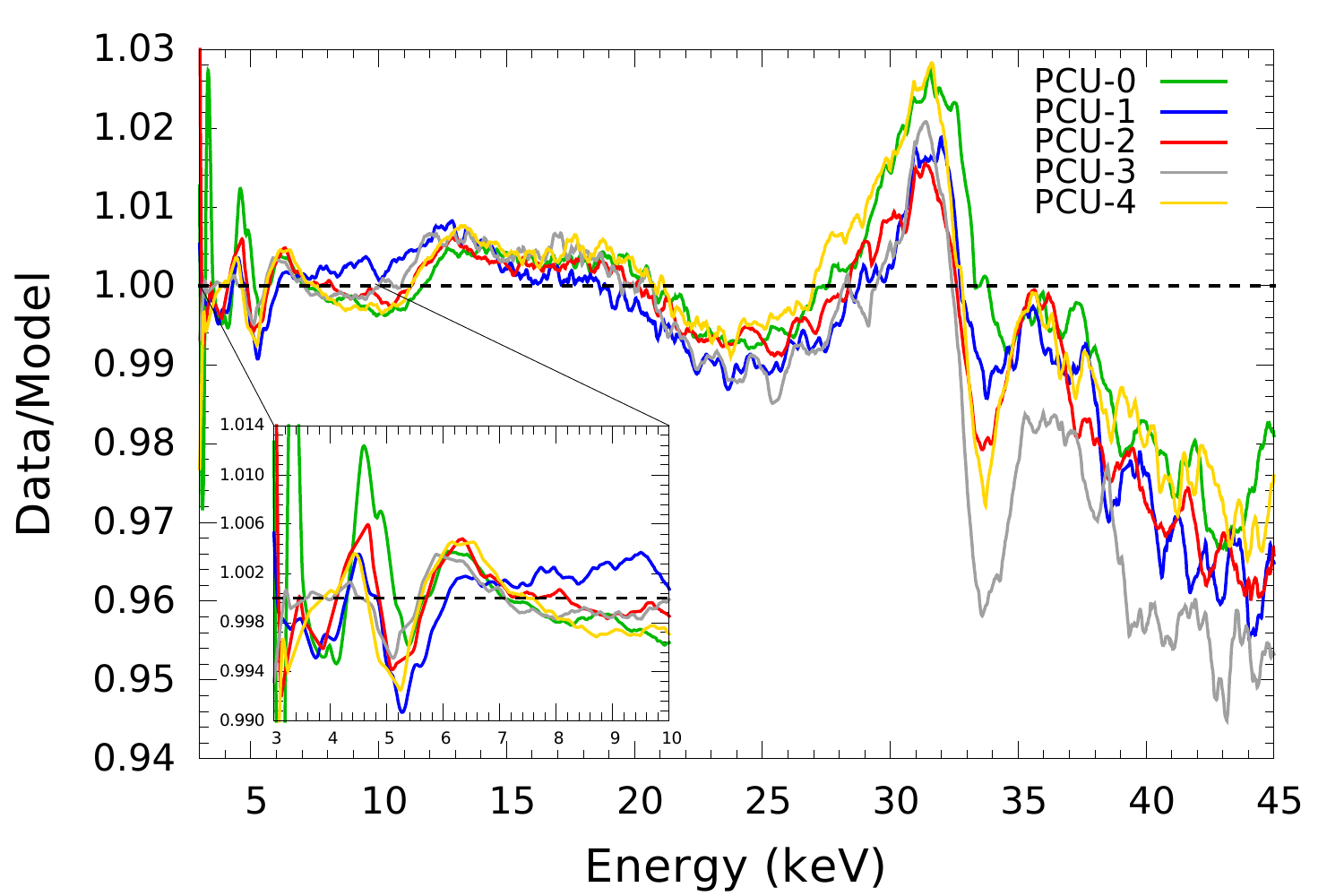}
\caption{Correction curves for all five PCUs using only observations
  taken during Gain Epochs~4--6. For details on the input spectra, see
  Table~\ref{tab:allpcu}.}
\label{fig:allpcu-ratio}
\end{center}
\end{figure}

As discussed in Section~\ref{sec:fits}, the channel boundaries change
with time. In addition, the energy-to-channel conversion is different
for each of the PCU detectors. Accordingly, one expects that if the
features in the ratio spectra are attributable to unmodeled instrumental
artifacts, they should differ between detectors. We have confirmed this
expectation by repeating the process described in Section~\ref{sec:corr}
for each of the PCUs. As before and for simplicity, we only show the
results of combining observations taken during Gain Epochs~4--6. An
overlay showing the correction curves for each detector is presented in
Figure~\ref{fig:allpcu-ratio}. Given that only a subset of the PCUs are
used in a given observation, each unit has a different available
exposure time and total counts. This information is summarized in
Table~\ref{tab:allpcu}.

%==================================================================================
%
\begin{deluxetable}{cccc}
\tabletypesize{\scriptsize}
\tablecaption{Details of the Crab observations for Gain Epochs~4--6
\label{tab:allpcu}}
\tablewidth{0pt}
\tablehead{
\colhead{PCU} & \colhead{No. of Spectra} & \colhead{Exposure Time} & \colhead{No. of Counts} \\
\colhead{   } & \colhead{              } & \colhead{          (s)} & \colhead{     ($10^9$)} \\
}
\startdata
0  & 373 & 447986 & $1.07$ \\
1  & 325 & 313936 & $0.73$ \\
2  & 417 & 483826 & $1.05$ \\
3  & 286 & 288336 & $0.64$ \\
4  & 329 & 316144 & $0.64$ \\
\enddata
\end{deluxetable}
%
%==================================================================================

As shown in Figure~\ref{fig:allpcu-ratio}, the correction curves of the five
PCUs are quite similar in several respects. Notably, the most prominent
features are present in all of the curves. For example, the narrow
absorption feature near 34 keV (presumably due to the Xe K-edge) and the
associated excess around 30--32 keV are pronounced for all the PCUs.
Also, all of the curves roll off at high energies, although the effect
is significantly stronger for PCU-3. Meanwhile, there are significant
differences: The corrections for PCU-1 deviates from the norm for energies
in the range $\sim7-20$~keV. Of the two narrow absorption features seen
in the correction for PCU-2 at $\approx 4.0$~keV and $\approx
5.5$~keV (right panel Figure~\ref{fig:ratio345}), only the latter
feature is common to all the PCUs (see the inset in
Figure~\ref{fig:allpcu-ratio}).

%==================================================================================
%
%==================================================================================

%
%==================================================================================
%
\section{Temporal Stability of the Correction Curves}\label{sec:temporal}

Given that the instrument response changes gradually as the detectors
age, and even suddenly, e.g., in the case of the loss of propane layers
of PCU-0 (May, 2000) and PCU-1 (December, 2006), we have studied the
stability of the correction over time. Here, we again discuss only our
results for PCU-2, while noting that we have checked and found quite
similar results for all the PCUs. Also, we consider only the stability
of the correction curve for Gain Epochs 4--6 because it covers 80\% of
the mission. To assess stability, we have produced three additional
correction curves by partitioning Epochs 4--6 with its 417 Crab
observations into three sequential time intervals, each containing 139
observations: (1) spectra obtained between 24 March 1999 and 26
September 2002 (MJD 51261--52543); (2) spectra obtained between 10
October 2002 through 17 November 2006 (MJD 52557--54056); and (3)
spectra collected during the final phase of the mission, after 18
November 2006 (MJD 54057).

\begin{figure}
\begin{center}
\includegraphics[scale=0.5,angle=0]{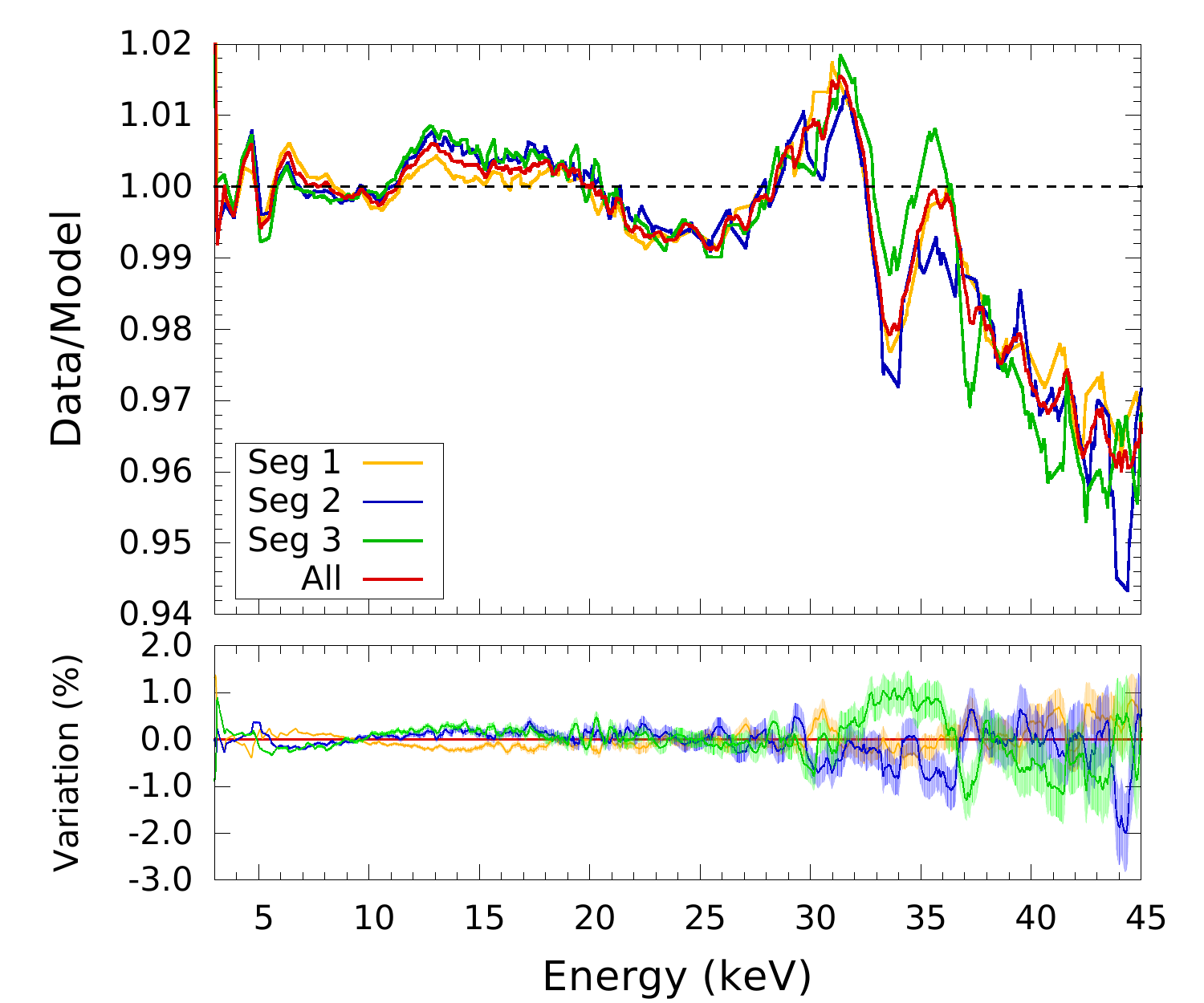}
\caption{Top: Correction curves for PCU-2 in three sequential time intervals
  covering Gain Epochs 4--6. For comparison, the full correction curve
  for the entire period is also shown (red line). Bottom: The three individual
  correction spectra divided by the full correction curve. The statistical
  error for each curve is indicated by the width of the shaded region.}
\label{fig:time-ratio}
\end{center}
\end{figure}

The correction curve for each time interval and an average over the
entirety are compared in the upper panel of
Figure~\ref{fig:time-ratio}. The lower panel of
Figure~\ref{fig:time-ratio} shows the percentage deviation of each
segment with respect to the total correction curve (also shown in
Figure~\ref{fig:ratio345}). The correction curves produced for the 
three time intervals are very close to each other for most of the 
channels, with the most pronounced differences ($\sim 3\%$) at high
energies where the statistical precision is relatively poor. We conclude
that the response of a given PCU is stable over a gain epoch, which
justifies using a single curve averaged over a gain epoch to correct a
spectrum of interest.

%==================================================================================
%
%==================================================================================

%
%==================================================================================
%
\section{Correction Curve for a Single Xenon Layer}\label{sec:layers}

\begin{figure}
\begin{center}
\includegraphics[scale=0.5,angle=0]{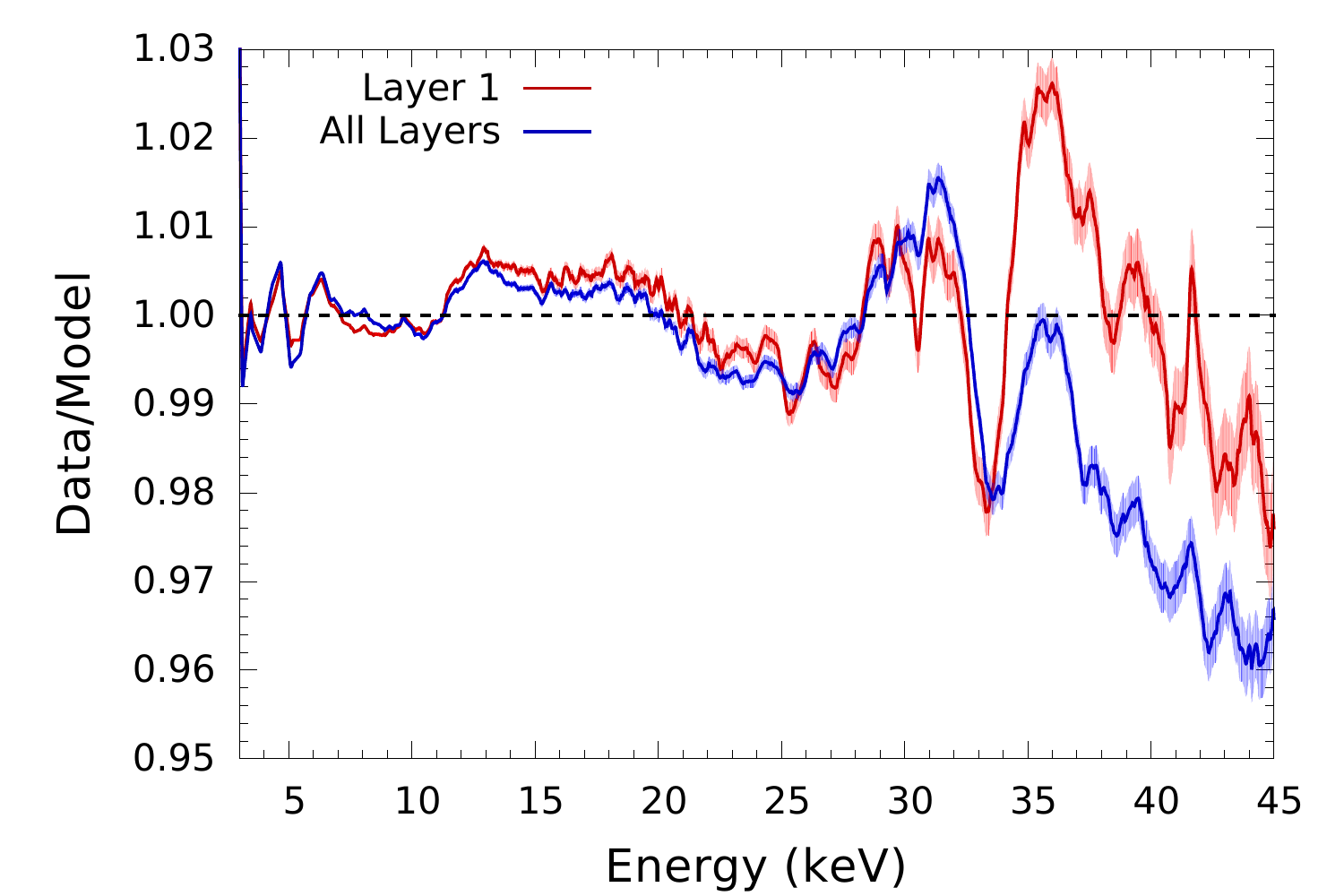}
\caption{Correction curves derived using the 417 PCU-2 Crab spectra from
Epochs~4--6. The blue spectrum was produced for the full 3-layer
detector (the same curve shown in the right panel of Figure~\ref{fig:ratio345}),
while the red curve was produced for the top layer only.
The lighter shaded region around each curve shows the level of the
statistical error.}
\label{fig:ratio-layers}
\end{center}
\end{figure}

Each PCU is a three-layered detector. In reducing PCA data, one can select data
either for all three layers combined or for the top layer only. All the results
that we have presented so far are for the full three-layer detector. Meanwhile,
we have also computed the single-layer correction for all the PCUs following
exactly the same procedures described in Sections~\ref{sec:fits} and
\ref{sec:corr}. The correction curve for PCU-2 for the top layer only is shown
in Figure~\ref{fig:ratio-layers}, where it is compared to the correction curve
for all three layers. This latter correction is identical to the one that
appears in the right panel of Figure~\ref{fig:ratio345}. Both PCU-2 corrections
were computed using the 417 Crab spectra collected during Gain Epochs 4--6.

There are significant differences between the two correction curves. Most
notably, the roll off at high energies, which is present in the 3-layer
correction of all of the PCUs (Figure~\ref{sec:allpcus}), is less pronounced in
the correction computed for a single layer; this is true for all the PCUs and
is illustrated for PCU--2 in Figure~\ref{fig:ratio-layers}. This difference
suggests that the roll off in the 3-layer correction curve discussed in
Section~\ref{sec:fits} is due to an error in the model of the detector
response, rather than to a break in the spectrum of the Crab. Because the
single-layer and 3-layer ratio spectra are distinctly different, we have
computed a total of 20 correction curves (2 detector configurations times 2
gain epochs times 5 PCUs).

%==================================================================================
%
%==================================================================================

%
%==================================================================================
%
\section{Discussion and Conclusions}\label{sec:disc}

A precise physical interpretation of the residual features seen in the
correction curves (Figures~\ref{fig:ratio345}, \ref{fig:allpcu-ratio}, and
\ref{fig:ratio-layers}) is beyond the scope of this paper. Nevertheless,
we identify and comment on three possible origins of these features,
namely, that they are: (i) intrinsic to the Crab; (ii)
imperfectly-modeled instrumental features due, e.g., to the Xe L- and
K-edges; or (iii) produced by inaccuracies in defining the channel
boundaries. The first two possibilities are expected to produce features
common to all the PCUs, while the latter is expected to produce narrow
features that differ among the PCUs because the gains of the individual
detectors differ.

The hypothesis (i) that the discrete features in the ratio spectra are
present in the spectrum of the Crab is disfavored because of our success
in correcting the spectra of GX~339--4, H1743--322, and XTE J1550--564;
furthermore, such features are not expected to be present in the
synchrotron spectrum of the Crab. However, as discussed in
Section~\ref{sec:fits}, the gradual roll off of the 3-layer correction
curve at energies $\gtrsim20$~keV may be attributable to a break in
the spectrum of the Crab.

Hypothesis (ii) is favored by the many remarkably similar features that
appear in the correction curves of the five PCUs
(Figure~\ref{fig:allpcu-ratio}), which strongly indicates that the
features are instrumental. Below $\sim7$~keV, however, there are narrow
features in the correction curves that vary from PCU to PCU (see
Section~\ref{sec:allpcus}); these discordant features favor hypothesis
(iii), indicating that the energies assigned to the channel boundaries
are inaccurate.
 
Regardless of the cause of the features, we have demonstrated that one
can use spectra of the Crab to routinely and significantly reduce
systematic error and improve the quality of PCA spectra; the results are
dramatic for high signal-to-noise data, with increases in sensitivity to
faint spectral features for bright sources by up to an order of
magnitude. As we have shown, the method works regardless of spectral
shape, e.g., whether the object spectrum is thermal and soft or whether
it is dominated by a hard power law. Furthermore, using a sophisticated
reflection model and a homogeneous sample of hard-state spectra of
GX~339--4, we found only modest changes in some of the parameters,
namely, those tied to the reflection features (such as inclination and
inner-disk radius). Meanwhile, the model parameters that determine the
continuum were unaffected.

In summary, by analyzing more than 1 billion counts at once we have
sensitively detected faint residual features in global spectra of the
Crab Nebula. As we have demonstrated, using the {\tt pcacorr} tool to
correct an arbitrary object spectrum with $\gtrsim10^6$ counts
significantly improves the quality of the fit. We have also
demonstrated that the inclusion of systematic uncertainty at the level
of 0.1\% is sufficient to achieve acceptable fits ($\chi_{\nu}^2\sim
1$; Figure~\ref{fig:chi2corr}), and we recommend routine use of 0.1\%
systematic error when analyzing PCA data that have been corrected with
the present tool. Following a period of further testing of {\tt
pcacorr} on a wider range of black-hole and neutron-star spectra, a
Python script that automates the correction of any PCA spectrum of
interest, along with a complete set of 20 correction curves (2 detector
configurations times 2 sets of gain epoch times 5 PCUs) will be made publicly
available at \url{http://hea-www.cfa.harvard.edu/~javier/pcacorr/}.
%
%
%==================================================================================
%
%
\acknowledgments JG and JEM acknowledge the support of NASA grant
NNX11AD08G. JFS has been supported by NASA Hubble Fellowship grant
HST-HF-51315.01. VG acknowledges support provided by NASA through the
Smithsonian Astrophysical Observatory (SAO) contract SV3-73016 to MIT
for support of the Chandra X-Ray Center (CXC) and Science Instruments;
CXC is operated by SAO for and on behalf of NASA under contract
NAS8-03060.  We thank the anonymous referee for a careful reading of our
paper, thoughtful criticisms and useful suggestions.  We also thank
J\"{o}rn Wilms, Keith Jahoda, Craig Markwardt, and Nikolai Shaposhnikov
for useful suggestions and valuable discussions.
%
%
%==============================================================================
%
%
\bibliographystyle{apj}
\bibliography{my-references}
%
%==================================================================================
%
%
%
\end{document}